\documentclass[showpacs,superscriptaddress,10pt,twocolumn,prb]{revtex4-1}
\usepackage{graphicx}
\usepackage{color}
\usepackage{dcolumn}

\begin{document}

% Use the \preprint command to place your local institutional report
% number in the upper righthand corner of the title page in preprint mode.
% Multiple \preprint commands are allowed.
% Use the 'preprintnumbers' class option to override journal defaults
% to display numbers if necessary
%\preprint{}

\title{Quasi-two-dimensional Dirac fermions and quantum magnetoresistance in LaAgBi$_2$}
\author{Kefeng Wang}
\affiliation{Condensed Matter Physics and Materials Science Department, Brookhaven National Laboratory, Upton, New York 11973, USA}
\author{D. Graf}
\affiliation{National High Magnetic Field Laboratory, Florida State University, Tallahassee, Florida 32306-4005, USA}
%\author{S. W. Tozer}
%\affiliation{National High Magnetic Field Laboratory, Florida State University, Tallahassee, Florida 32306-4005, USA}
\author{C. Petrovic}
\affiliation{Condensed Matter Physics and Materials Science Department, Brookhaven National Laboratory, Upton, New York 11973, USA}

\date{\today}

\begin{abstract}
We report quasi-two-dimensional Dirac fermions and quantum magnetoresistance in LaAgBi$_2$. The band structure shows several narrow bands with nearly linear energy dispersion and Dirac-cone-like points at the Fermi level. The quantum oscillation experiments revealed one quasi-two-dimensional Fermi pocket and another complex pocket with small cyclotron resonant mass. The in-plane transverse magnetoresistance exhibits a crossover at a critical field $B^*$ from semiclassical weak-field $B^2$ dependence to the high-field unsaturated linear magnetoresistance which is attributed to the quantum limit of the Dirac fermions. Our results suggest the existence of quasi 2D Dirac fermions in rare-earth based layered compounds with two-dimensional double-sized Bi square nets, similar to (Ca,Sr)MnBi$_{2}$, irrespective of magnetic order.
\end{abstract}
\pacs{72.20.My,72.80.-r,75.47.Np}

\maketitle

\section{Introduction}
Dirac fermions with linear energy-momentum dispersion and corresponding Dirac cone states have been observed in two-dimensional graphene \cite{graphene1,graphene2} and the surface of topological insulators (TI).\cite{ti1,ti2} It is believed that the two bands with the opposite (pseudo)spins cross each other without hybridization giving the linear energy dispersion. Unlike the conventional electron gas with parabolic energy dispersion where Landau levels (LLs) are equidistant,\cite{metal} the distance between the lowest and $1^{st}$ LLs of Dirac fermions in magnetic field is very large. So the quantum limit where all of the carriers occupy only the lowest LL is easily realized even in moderate fields.\cite{berry} Consequently some quantum transport phenomena such as quantum Hall effect and large linear magnetoresistance (MR) could be observed in the regular magnetic field in Dirac fermion system.\cite{berry,qt1,qt2,qt3,qt4} Thus, Dirac materials are now one of the central topics of condensed matter physics.

In addition to nanoengineered or surface materials such as graphene and TIs, the Dirac fermions and Dirac nodes were observed or predicted in bulk crystals of iron-based superconductor parent material BaFe$_2$As$_2$,\cite{qt3,qt4,iron1,iron2} (Sr/Ca)MnBi$_2$ bismuth based layered magnetic compounds\cite{srmnbi21,srmnbi22,srmnbi23,srmnbi24} and in a molecular organic conductor $\alpha-$(BEDT-TTF)$_2$I$_3$.\cite{organic} Among them, SrMnBi$_2$ contains alternatively stacked two MnBi$_4$ tetrahedron layers and a two-dimensional (2D) Bi square net separated by Ca atoms along the $c$-axis. The linear energy dispersion originates from the crossing of two Bi $6p_{x,y}$ bands in the double-sized Bi square nets corresponding to the double-sized N$\acute{e}e$l-type antiferromagnetic (AFM) Mn unit cell.\cite{srmnbi21} However, in a SrMnBi$_2$-type structure, the nonmangetic unit cell still contains two Bi atoms in the Bi square net due to the occupation of Sr atoms.\cite{lattice} In previous report,\cite{srmnbi21} all the analysis is based on the electronic structure of SrMnBi$_2$ in antiferromagnetic state. Hence, the effect of AFM order within MnBi$_4$ layers in the formation of Dirac nodes and whether any unit cell symmetry with double-sized Bi unit cell can host Dirac fermions is not completely clear. Besides, the engineering of Dirac states is of the great interest. With the change of band parameters in deformed graphene, Dirac points may merge or be completely removed.\cite{engineer} Similarly in bulk crystals, hopping terms and the band parameters can be tuned by changing the lattice parameters, hybridization or the space group of the crystallized structure. It was reported that SrMnBi$_2$ and CaMnBi$_2$ host Dirac dispersion of different nature. Even though the conduction and valence bands touch at the Dirac point in both materials, the details are different due to the different stacking of nearby alkaline earth atoms and the different hybridization. For SrMnBi$_2$, the zero-energy gap is found only at a specific point, while it is found along the continuous line in the momentum space for CaMnBi$_2$.\cite{SrCaMnBi2} Hence exploring new bulk compounds with similar structure to SrMnBi$_2$ may provide more profound comprehension of Dirac band crossing mechanisms in bulk crystal.

Here we report quasi-2D Dirac fermions in the LaAgBi$_2$ single crystal which has similar crystal lattice structure with CaMnBi$_2$, but without magnetic ions.\cite{laagbi1} The band structure shows several narrow bands with nearly linear energy dispersion and Dirac-cone-like points at the Fermi level. The quantum oscillation experiments show one quasi-two-dimensional Fermi pocket and another very complex electron pocket with small cyclotron resonant mass. The in-plane transverse magnetoresistance exhibits a crossover at a critical field $B^*$ from semiclassical weak-field $B^2$ dependence to the high-field unsaturated linear magnetoresistance due to the quantum limit of the Dirac fermions. The temperature dependence of $B^*$ satisfies quadratic behavior, which is attributed to the splitting of linear energy dispersion in high field. Our results demonstrate that Dirac fermions in bulk crystals can also be found in the absence of magnetic order and imply possible universal existence of two dimensional Dirac fermions in layered structure compounds with two-dimensional double-sized Bi square nets.

\section{Experimental}

Single crystals of LaAgBi$_2$ were grown using a high-temperature self-flux method.\cite{laagbi1} The resultant crystals are plate-like. X-ray diffraction (XRD) data were taken with Cu K$_{\alpha}$ ($\lambda=0.15418$ nm) radiation of Rigaku Miniflex powder diffractometer. Electrical transport measurements up to 9 T were conducted in Quantum Design PPMS-9 with conventional four-wire method. In the in-plane measurements, the crystal was mounted on a rotating stage such that the tilt angle $\theta$ between sample surface ($ab$-plane) and the magnetic field can be continuously changed, with currents flowing in the $ab$-plane perpendicular to magnetic field. The de Haas-van Alphen (dHvA) oscillation experiments were performed at National High Magnetic Field Laboratory, Tallahassee. The crystals were mounted onto miniature Seiko piezoresistive cantilevers which were installed on a rotating platform. The field direction can be changed continuously between parallel and perpendicular to the $c$-axis of the crystal. First principle electronic structure calculation were performed using experimental lattice parameters within the full-potential linearized augmented plane wave (LAPW) method ~\cite{wien2k1} implemented in WIEN2k package.\cite{wien2k2} The general gradient approximation (GGA) of Perdew \textit{et al}., was used for exchange-correlation potential.\cite{gga} The LAPW sphere radius were set to 2.5 Bohr for all atoms, and the converged basis corresponding to $R_{min}k_{max}=7$ with additional local orbital were used where $R_{min}$ is the minimum LAPW sphere radius and $k_{max}$ is the plane wave cutoff. Spin-orbit coupling for all elements were took into account by a second-variational method with the scalar-relativistic orbitals as basis which was implemented in WIEN2k.

\section{Results and discussions}

Fig. 1(a) shows the powder XRD pattern of flux grown LaAgBi$_2$ crystals, which were fitted by RIETICA software.\cite{rietica} All reflections can be indexed in the P4/nmm space group. The determined lattice parameters are $a=b=0.4582(8)$ nm and $c=1.062(4)$ nm, in agreement with the published data.\cite{laagbi1,laagbi2} The basal plane of a cleaved crystal is the crystallographic $ab$-plane where the 2D Bi layers (Bi2, the red balls in Fig. 1(b)) are located. Contrary to CaMnBi$_2$, the adjacent 2D Bi layers along $c$-axis are separated by La atoms and AgBi layers without magnetic ions (Fig. 1(b)). This makes LaAgBi$_2$ an ideal system to clarify the role of magnetic ions in the formation of Dirac fermions in CaMnBi$_2$ and other similar compounds.

Fig. 1(c) shows the first-principle band structure without spin-orbit coupling. Fig 1(d), Fig. 2(a) and Fig. 2(b-d) show the band structure and the density of states (DOS), and Fermi surfaces of LaAgBi$_2$ with spin-orbit coupling, respectively. The band structure in Fig. 1 clearly shows several narrow linear bands. More interesting, probably at the Fermi level, there are two Dirac-cone-like points along $\Gamma-M$, $R-Z$ directions and at $X$ points in the Brillouin zone (red circles in Fig. 1(d)). Compared to the band structure without spin-orbit interaction in Fig. 1(c), the spin-orbit coupling induces the gap at the Dirac-cone-like points and a lowering of the Fermi level around 20 meV in Fig. 1(d). Besides these, the band structure with and without spin-orbit coupling looks similar and the essential features of the FS's remain almost the same In Fig. 2(a), the Fermi level is located at the edge of the gap, and the main peaks of DOS from La, Ag and Bi1 are located far below the Fermi level.  The conducting electrons in LaAgBi$_2$ are mainly due to 5p from Bi2, while there is little contribution from other atoms as shown in Fig. 2(a). So the linear bands and the Dirac-cone like points at the Fermi level mainly originate from Bi bands. In (Sr/Ca)MnBi$_2$, the antiferromagnetic order of Mn moments doubles the unit cell. Consequently two Bi $6p_{x,y}$ bands in the double-sized Bi square nets cross each other without significant hybridization and form the linear bands and Dirac-cone like points.\cite{srmnbi21,srmnbi22} There are no magnetic ions in LaAgBi$_{2}$, but there are still two Bi2 atoms per unit cell because of the occupation of La ions (one above and another below the Bi2 layer), as shown in Fig. 1(b). This will lead to the folding of the dispersive $p$ orbital of Bi2. The two $p_{x,y}$ bands from two Bi2 atoms cross each other at a single point and then form the nearly linear band and Dirac-cone-like point around the Fermi level (Fig. 1(c)). Correspondingly the Fermi pockets along $\Gamma$-M (Fig. 2(d)), and the one along $R-Z$ directions and at $X$ points (Fig. 2(b)) host Dirac fermions.

It is important to compare the similarity/difference of the Dirac dispersion between LaAgBi$_2$ and (Sr/Ca)MnBi$_2$.  In (Sr/Ca)MnBi$_2$, conduction and valence bands touch at the Dirac point in both materials, but the details are different due to the different stacking of nearby alkaline earth atoms and the different hybridization. For SrMnBi$_2$, the zero-energy gap is found only at a specific point, while it is found along the continuous line in the momentum space for CaMnBi$_2$.\cite{SrCaMnBi2} Since the crystal structure of LaAgBi$_2$ is identical to CaMnBi$_2$, there is also continuous zero-energy gap line in LaAgBi$_2$ which is different from SrMnBi$_2$. But due to the different valence and hybridization between La and Ca ions as well as the nomagnetic Ag ions, some difference is expected in Fermi surfaces and Dirac dispersions.  In CaMnBi$_2$, the Dirac pocket (electrons) is only observed along $\Gamma$-X line,\cite{SrCaMnBi2} but in LaAgBi$_2$ there are Dirac electron pockets along $\Gamma$-X and $\Gamma-M$ lines (Fig. 1 and Fig. 2). In addition, the magnetic order of Mn ions is important to expel the states from the Fermi level, in contrast several regular parabolic bands cross the Fermi level in LaAgBi$_2$ (Fig. 1(c)). These observations indicate that the magnetic order in (Sr/Ca)MnBi$_2$ is not the key ingredient in the Dirac cone formation mechanism in 2D Bi layers and points out a direction for the search of new Dirac materials. Nevertheless, the magnetic order or electronic correlation can still remove other parabolic bands from the Fermi level.

%%%%%%%%%%%%%%%%%%%%%%% Figure 1 %%%%%%%%%%%%%%%%%%%%%%%%%%%%%%%%%%
\begin{figure}[tbp]
\includegraphics[scale=0.42]{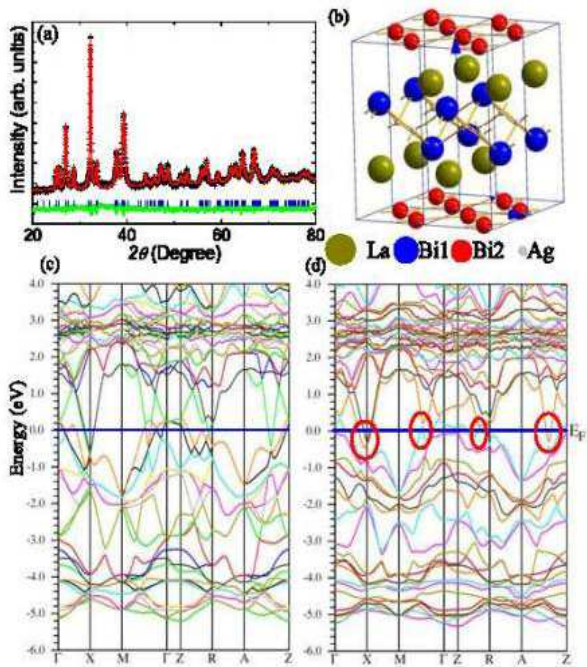}
\caption{(color online) (a) Powder XRD patterns and structural refinement results. The data were shown by ($+$) , and the fit is given by the heavy solid line. The difference curve (the light solid line) is offset and the segments indicate the observed peaks. (b) The crystal structure of LaAgBi$_2$. Atoms are distinguished by their size starting from La (largest) to Ag (smallest). Bi atoms in 2D square nets (Bi2) are somewhat smaller than Bi atoms in AgSb4 tetrahedra (Bi1). (c) and (d) The band structure for LaAgBi$_2$ with (c) and without (d) spin-orbit coupling effect. The different bands were indicated by different color. The line at Energy=0 indicates the position of Fermi level. The red circles denote the position of the Dirac-cone-like points close to the Fermi level.}
\end{figure}
%%%%%%%%%%%%%%%%%%%%%%%%%%%%%%%%%%%%%%%%%%%%%%%%%%%%%%%%%%%%%%%%%%%%

%%%%%%%%%%%%%%%%%%%%%%% Figure 2 %%%%%%%%%%%%%%%%%%%%%%%%%%%%%%%%%%
\begin{figure}[tbp]
\includegraphics[scale=0.4] {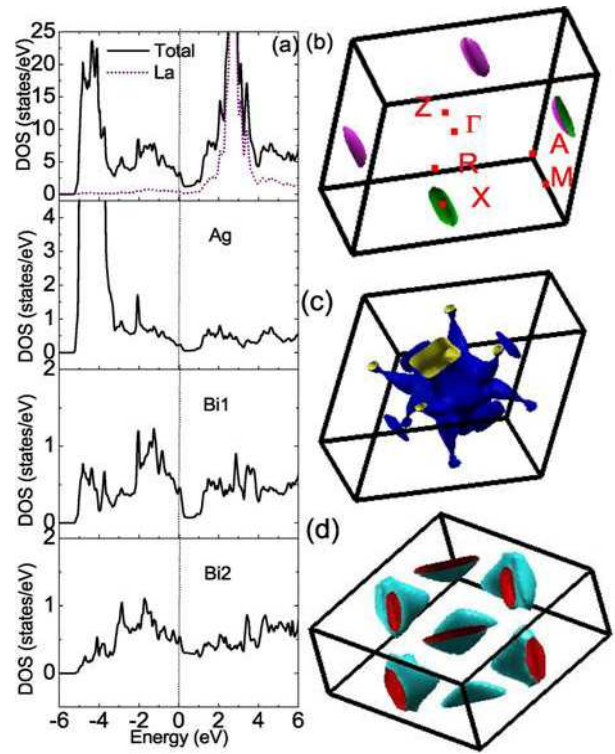}
\caption{(color online) (a) The total density of states (black line) and local DOS from La (dot line) (upper panel), Ag (the second panel),Bi1 (the third panel) and Bi2 (bottom panel) in LaAgBi$_2$. The dotted line indicates the position of the Fermi energy. (b,c,d) The shape of three different Fermi pockets of LaAgBi$_2$.}
\end{figure}
%%%%%%%%%%%%%%%%%%%%%%%%%%%%%%%%%%%%%%%%%%%%%%%%%%%%%%%%%%%%%%%%%%%%

%%%%%%%%%%%%%%%%%%%%%%% Figure 3 %%%%%%%%%%%%%%%%%%%%%%%%%%%%%%%%%%
\begin{figure}[tbp]
\includegraphics[scale=0.4] {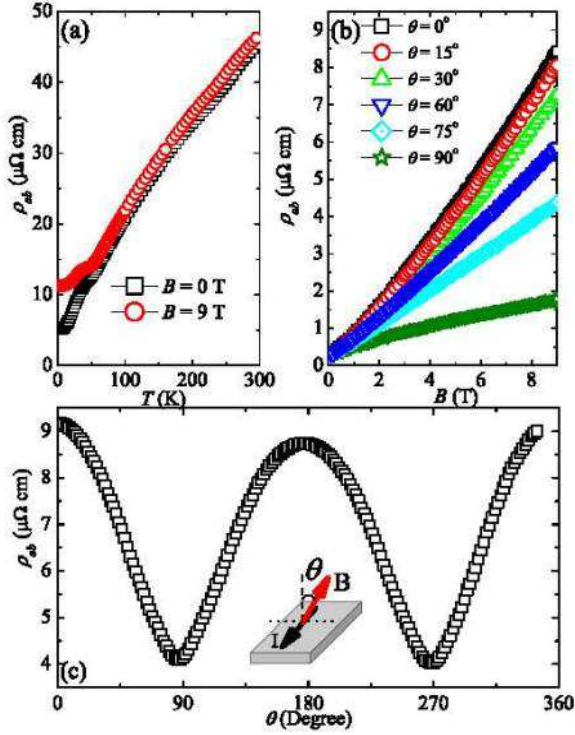}
\caption{(color online) (a) In-plane resistivity $\rho_{ab}(T)$ of LaAgBi$_2$ single crystal in 0 T and 9 T magnetic field. (b) In-plane Resistivity $\protect\rho$ vs. magnetic field $B$ of LaAgBi$_2$ crystal with different tilt angle $\protect\theta$ between magnetic field and sample surface ($ab$-plane) at 2 K. (c) In-plane resistivity $\rho$ vs. the tilt angle $\protect\theta$ from $0^o$ to $360^o$ at $B$ = 9 T and $T$ = 2 K. Inset shows the configuration of the measurement.}
\end{figure}
%%%%%%%%%%%%%%%%%%%%%%%%%%%%%%%%%%%%%%%%%%%%%%%%%%%%%%%%%%%%%%%%%%%%

Linear bands and Dirac fermions have considerable effects on the transport properties of materials. The in-plane resistivity $\rho_{ab}$ of LaAgBi$_2$ single crystal is metallic in the whole temperature range with a significant anomaly at $\sim 30$ K (Fig 3(a)). Similar behavior was observed in LaAgSb$_2$ which was attributed to the possible charge density wave (CDW) order and possibly implies the same order/transition in LaAgBi$_2$. The external magnetic field significantly enhances the resistivity below 30 K, but has little influence on the transport behavior above 30 K (Fig. 3(a)). The magnetoresistance of LaAgBi$_2$ shows significant dependence on the field direction (Fig 3(b,c)). The crystal was mounted on a rotating stage such that the tilt angle $\theta$ between the crystal surface ($ab$-plane) and the magnetic field can be continuously changed with currents flowing in the $ab$-plane perpendicular to magnetic field, as shown in the inset of Fig. 3(c).  Angular dependent magnetoresistance $\rho(B,\theta)$ at $T\sim 2$ K is shown in Fig. 3(b) and (c). When $B$ is parallel to the $c$-axis ($\theta=0^{o}, 180^{o}$), the MR is maximized and is linear in field for high fields. With increase in the tilt angle $\theta$, the MR gradually decreases and becomes nearly negligible for $B$ in the $ab$-plane ($\theta=90^o$), as shown in Fig. 3(b). Angular dependent resistivity in $B=9$ T and $T=2$ K shows wide maximum when the field is parallel to the $c$-axis ($\theta=0^o, 180^o$), and sharper minimum around $\theta=90^o, 270^o$ (Fig. 3(c)). Hence, the Fermi surface of LaAgBi$_2$ is highly anisotropic along $ab$-plane and $c$-axis.

%%%%%%%%%%%%%%%%%%%%%%% Figure 4 %%%%%%%%%%%%%%%%%%%%%%%%%%%%%%%%%%
\begin{figure}[tbp]
\includegraphics[scale=1] {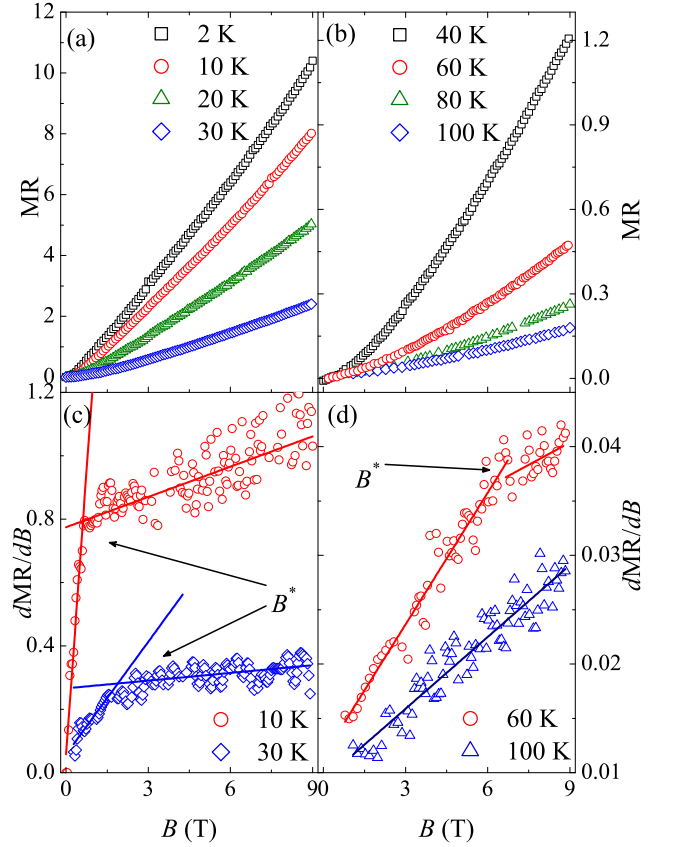}
\caption{(color online) (a) and (b) The magnetic field ($B$) dependence of the in-plane magnetoresistance MR at different
temperatures. (c) and (d) The field derivative of in-plane MR, $d$MR$/dB$, as a function of field (B) at different temperature
respectively. The solid lines in high field regions were fitting results using MR $=A_1B+O(B^2)$ and the lines in low field regions using MR $=A_2B^2$.}
\end{figure}
%%%%%%%%%%%%%%%%%%%%%%%%%%%%%%%%%%%%%%%%%%%%%%%%%%%%%%%%%%%%%%%%%%%%

LaAgBi$_2$ exhibits very large linear magnetoresistance. At 2 K, the MR is linear in the high field region and reaches $\sim 1200\%$ in 9 T field (Fig. 4 (a)). This linear behavior extends to a very low crossover fields $B^*$ where the MR naturally reduces to a weak-field semiclassical quadratic response. In order to extrapolate the crossover field $B^*$, we plot  the field derivative of MR, $d$MR$/dB$, in Fig. 4(b) and (c). In the low field range ($B<$1 T at 2 K), $d$MR$/dB$ is proportional to $B$ (as shown by lines in low-field regions), indicating the semiclassical MR $\sim A_2B^2$. But above a characteristic field $B^*$, $d$MR$/dB$ deviates from the
semiclassical behavior and saturates to a much reduced slope (as shown by lines in the high-field region). This indicates that the MR for $B>B^*$ is dominated
by a linear field dependence plus a very small quadratic term (MR$ =A_1B+O(B^2)$). With increasing temperature, the field range where linear MR appears shrinks and MR decreases. Ultimately we cannot observe any linear MR below 9 T and above 100 K. The cross-over field $B^*$ is defined as the value where the fitting lines cross each other. Below 9 T and 50 K, the evolution of $B^*$ with temperature is parabolic (Fig. 5(a)).

%%%%%%%%%%%%%%%%%%%%%%% Figure 5 %%%%%%%%%%%%%%%%%%%%%%%%%%%%%%%%%%
\begin{figure}[tbp]
\includegraphics [scale=0.8]{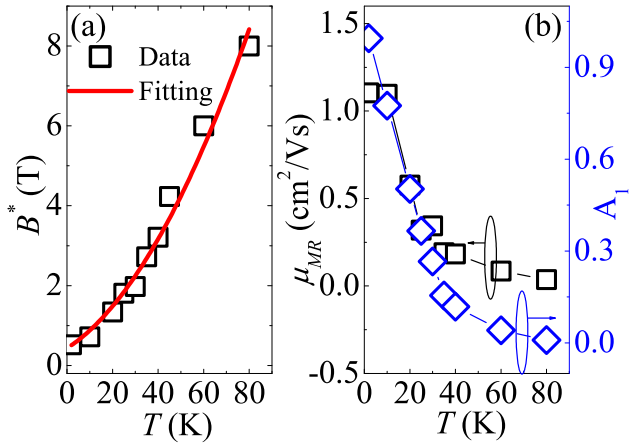}
\caption{(color online) (a)Temperature dependence of the critical field $B^*$ (black
squares) The solid line is the fitting results of $B^*$ using $B^*=\frac{1}{2e\hbar
v_F^2}(E_F+k_BT)^2$. (b) The effective MR mobility $\mu_{MR}$ (black squares) extracted from the weak-field MR and the fitting coefficient $A_1$ for the linear term in MR.}
\end{figure}
%%%%%%%%%%%%%%%%%%%%%%%%%%%%%%%%%%%%%%%%%%%%%%%%%%%%%%%%%%%%%%%%%%%%

The linear MR which evidently deviates from semiclassical transport behavior in metal has been observed in bulk crystals of Ag$_{2-\delta}$(Te/Se),\cite{agte1,agte2} Bi$_2$Te$_3$,\cite{qt1,qt2,qt4} BaFe$_2$As$_2$\cite{qt3} and (Sr/Ca)MnBi$_2$.\cite{srmnbi21,srmnbi22,srmnbi23,srmnbi24} Among them, Ag$_{2-\delta}$(Te/Se) and Bi$_2$Te$_3$ were found to be topological insulators hosting topological protected Dirac surface states,\cite{fangzhong,ti1,ti2} while BaFe$_2$As$_2$ and (Sr/Ca)MnBi$_2$ host linear bands in the bulk.\cite{qt3,iron1,iron2,srmnbi21} The energy splitting between the lowest and $1^{st}$ LLs of Dirac fermions can be described by $\triangle_{LL}=\pm v_F\sqrt{2e\hbar B}$ where $v_F$ is the Fermi velocity.\cite{LL1,LL2,qt1,qt2} In the quantum limit $\triangle_{LL}$ is larger than both the Fermi energy $E_F$ and the thermal fluctuations $k_BT$. All carriers occupy the lowest Landau level and therefore the quantum transport with linear magnetoresistance dominates the conduction. The critical field $B^*$ above which the quantum limit is satisfied at specific temperature $T$ is $B^*=\frac{1}{2e\hbar v_F^2} (E_F+k_BT)^2$.\cite{qt3} The $B^*(T)$ in LaAgBi$_2$ can be well fitted by the above equation, as shown by the solid line in Fig. 5(a). This confirms the existence of Dirac fermion states in LaAgBi$_2$. In a multiband system with dominant Dirac states and conventional parabolic-band carriers (including electrons and holes), the coefficient of the low-field semiclassical $B^2$ quadratic term, $A_2$, is related to the effective MR mobility $\sqrt{A_2}=\frac{\sqrt{\sigma_e\sigma_h}}{\sigma_e+\sigma_h}(\mu_e+\mu_h)=\mu_{MR}$ (where $\sigma_e, \sigma_h, \mu_e, \mu_h$ are the effective electron and hole conductivity and mobility in zero field respectively). The effective MR is smaller than the average mobility of carriers $\mu_{ave}=\frac{\mu_e+\mu_h}{2}$ and gives an estimate of the lower bound.\cite{qt3,qt4} Fig. 5(b) shows the dependence of $\mu_{MR}$ on the temperature. At 2 K, the value of $\mu_{MR}$ is about 1200 cm$^2$/Vs in LaAgBi$_2$ which is larger than the values in conventional metals. With increasing temperature, the linear MR, the linear term coefficient $A_1$ and $\mu_{MR}$ are suppressed due to the temperature smearing of the Landau level splitting.

%%%%%%%%%%%%%%%%%%%%%%% Figure 6 %%%%%%%%%%%%%%%%%%%%%%%%%%%%%%%%%%
\begin{figure}[tbp]
\includegraphics [scale=1]{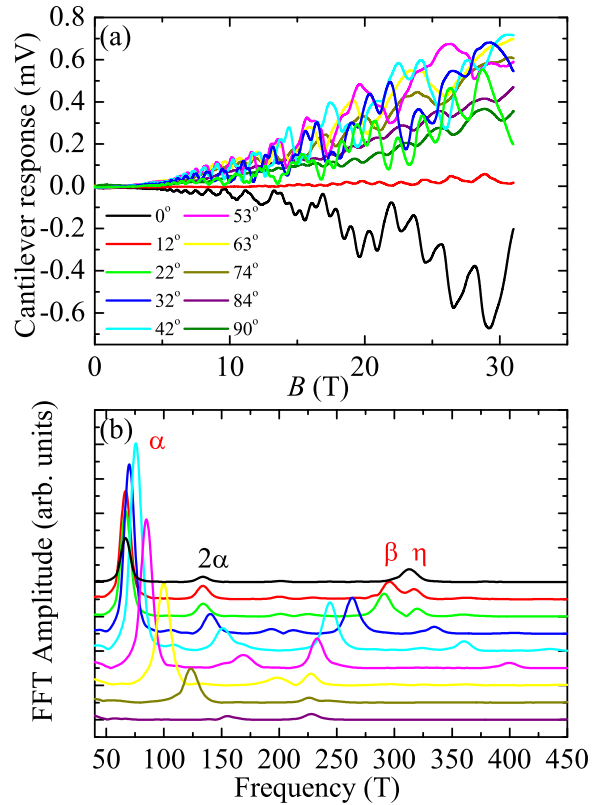}
\caption{(color online) (a) Quantum oscillation of LaAgBi$_2$ observed with cantilever as a function of the magnetic field ($B$) with different tilting angle. (b) The Fourier transform spectrum of the SdH oscillation. For the peak labels see the main text.}
\end{figure}
%%%%%%%%%%%%%%%%%%%%%%%%%%%%%%%%%%%%%%%%%%%%%%%%%%%%%%%%%%%%%%%%%%%%

In order to further clarify the electronic structure and Dirac fermions in LaAgBi$_2$, we performed first principle Fermi surface calculations and dHvA oscillation measurement on LaAgBi$_2$ single crystal. Fig. 2(b-d) shows the topology of the theoretical Fermi surfaces of three different pockets for LaAgBi$_2$, respectively. Centered at $X$ point, there are very small ellipsoid electronic pockets with the long axis along $\Gamma-Z$ directions, corresponding to the Dirac-cone-like point at $X$ point in band structure.This Fermi pocket predicts the presence of a dHvA oscillation frequency of about 82 T. At the center of the Brillouin zone, there is a big hollow cylindrical hole pocket.This hole pockets is nearly ten times larger than previous ellipsoid pocket and predicts a frequency of $\sim 1000$ T. These two pockets are nearly isotropic in $ab$-plane but different along $c$-axis. In addition, there is another complex electron pocket along diagonal direction. This pocket is highly anisotropic along three axis, corresponding to the anisotropic point in $\Gamma-M$ direction in band structure. This complex pocket will give two similar frequencies of 370 T and 320 T. It is of interest to note that the theoretical hole and electron Fermi surfaces are compensated, similar to semimetals.

The quantum oscillation provide a direct probe of the Fermi surface. In metals, quantum oscillations correspond to successive emptying of LLs by the magnetic field and the oscillation frequency is related to the cross section area of the Fermi surface $S_F$ by the Onsager relation $F=(\hbar/2\pi e)S_F$.\cite{osc1,qt1,qt2} From the temperature evolution of the oscillation amplitude, one can deduce the effective cyclotron resonant mass by the fitting using Lifshitz-Kosevitch formula.\cite{osc1} From the evolution of these frequencies as a function of the magnetic field orientation ($\theta$) and temperature, one can construct a detailed picture of the size and shape of the Fermi surface. Fig. 6(a) and (b) show the typical dHvA oscillations and the Fourier transform (FFT) spectrum of the oscillations for LaAgBi$_2$ single crystal with different magnetic field direction. When the magnetic field is close to perpendicular to the $ab$ plane ($\theta$ close to $0^o$), the signal exhibits significant oscillation (Fig. 6(a)). The FFT spectrum of the oscillations (Fig. 6(b)) exhibit two peaks at $\sim 67$ and 300 T which are labeled as $\alpha$ and $\beta$. There is another peak at 135 T which corresponds to the double frequency of peak $\alpha$. With increasing $\theta$, oscillations are weaker and the peak $\alpha$ shifts to higher frequency. More interestingly, the peak $\beta$ splits to two peaks (labeled as $\beta$ and $\eta$). One of them ($\beta$) shifts to lower frequency but another ($\eta$) shifts toward higher frequency with increasing angle.

%%%%%%%%%%%%%%%%%%%%%%% Figure 7 %%%%%%%%%%%%%%%%%%%%%%%%%%%%%%%%%%
\begin{figure}[tbp]
\includegraphics [scale=0.4]{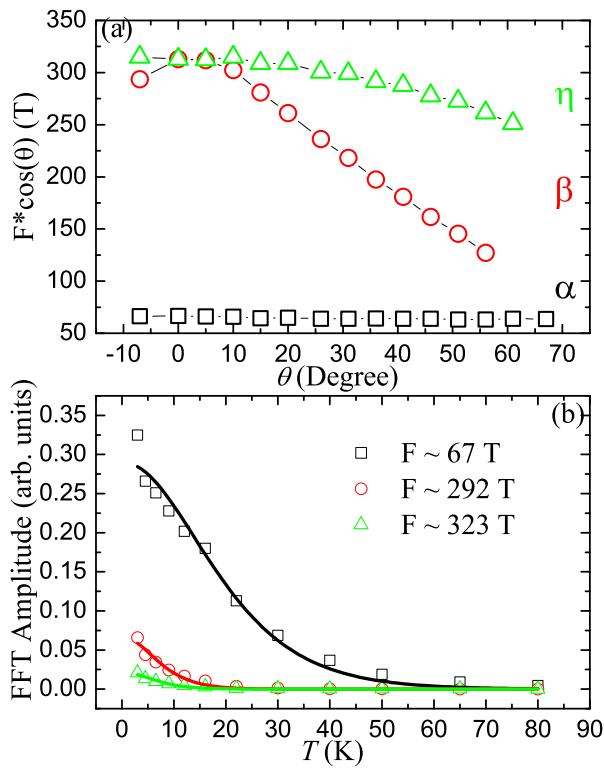}
\caption{(color online) (a) The evolution of the de Hass-van Alphen oscillation frequencies $F$ plotted as $F*\cos(\theta)$ with the magnetic field angle $\theta$. (b) Temperature dependence of the oscillation amplitudes (Osc. Amp.) in quantum oscillation using cantilevert. The discrete symbols are the experimental results and the solid lines are the fitting results giving cyclotron mass.}
\end{figure}
%%%%%%%%%%%%%%%%%%%%%%%%%%%%%%%%%%%%%%%%%%%%%%%%%%%%%%%%%%%%%%%%%%%%

The detailed angular dependence of the three FFT frequencies in the oscillation are shown in Fig. 7(a). For a 2D FS (a cylinder), the cross section has $S_F(\theta)=S_0/|\cos(\theta)|$ angular dependence and the oscillation frequencies should be inversely proportional to $|cos(\theta)|$. For a sphere Fermi pocket, the cross section for all magnetic field direction is a constant and the oscillation frequencies should not be angular dependent.\cite{qt1,osc1} In Fig. 7(a), the angular dependence of the oscillation frequency $\alpha$ ($\sim 67$ T at $\theta=0$) multiplied by $\cos{\theta}$ (open squares) does not show significant dependence on the field angle $\theta$, indicating a quasi-2D Fermi pocket. The temperature evolution of the oscillation amplitude gives a cyclotron mass $m\sim 0.056m_e$ where $m_e$ is the bare electron mass (Fig. 7(b)) and similar to previous observation.\cite{laagbi1} The oscillation frequency and mass is close to the ellipsoid electron pocket at $X$ point. Other two pockets $\beta$ and $\eta$ should not come from the hollow cylinder hole pocket at the center of the Brillouin zone, because the hole pocket is nearly one hundred times bigger than the ellipsoid electron pocket but the oscillation frequencies for $\beta$ and $\eta$ is only ten times bigger the $\alpha$. Moreover, these two pockets show similar effective mass ($m_{\beta}\sim 0.14(2)m_e, m_{\eta}\sim0.16(5)m_e$) (Fig. 7(b)). These two oscillation frequencies should correspond to the highly anisotropic pockets locating along the edge and diagonal directions in Brillouin zone (Fig. 2(d)) since the oscillation frequencies were close to the theoretical values. but in our in-plane measurement, the magnetic field is along [100] direction. Since we cannot change the in-plane field direction, we can not distinguish these two pockets from the angular dependent oscillation frequency in present measurement. For pocket $\eta$, the quantum oscillation frequency, $F\times\cos{\theta}$ decreases a little bit with increasing $\theta$, but its change is much smaller when compared to pocket $\beta$. This is consistent with the calculated Fermi surface (Fig. 2(d)), which shows that the dispersion along $k_z$ direction is much smaller than the value along $k_x$ and $k_y$ directions. Besides that, the Dirac fermions with linear bands have much larger mobility and Fermi velocity than the regular carriers, and will dominate the transport properties. Hence, the angular dependent magnetoresistance (Fig. 3(b) and (c)) should come from the quasi-two-dimensional Dirac Fermi pockets.

Our results demonstrate the possible universal existence of two dimensional Dirac fermions in layered structure compounds with two-dimensional Bi square nets, irrespective of magnetic order. So it is important to study the relationship between the CDW transition and the Dirac fermions. For the systems with charge density wave or spin density wave, the phase transition often induces band-folding of some Fermi surface sections. CDW was found in LaAgSb$_2$ and LaSb$_2$, where calculated Fermi surfaces without CDW transition agree very well with low-temperature quantum oscillation results in CDW state.\cite{oscillation1,oscillation2,oscillation3} In LaAgBi$_2$, the observed three frequencies are consistent with the calculated Fermi surfaces along $\Gamma$-M and $\Gamma$-X directions (Fig. 2(b) and (d)) after a moderate energy shift ($\sim$ 20 meV), but the large hole pockets centered at $\Gamma$ point is absent in the dHvA oscillation, which is similar to the results in LaAgSb2.\cite{oscillation1,oscillation3} Hence, the CDW transition most likely smears out the large hole pocket and induces the shift of the energy of other three pockets which host Dirac fermions. The x-ray scattering experiments in LaAgSb$_2$ revealed that the nesting of the Fermi surfaces responsible for the two CDWs happens in band 1 centered at $\Gamma$ point and band 3 extending throughout the zone with the vertices of the squares at the X point. The Fermi surfaces of LaAgBi$_2$ should be similar to these of LaAgSb$_2$ because of the similar structure. So it could be expected that the nesting happens in the band centered at $\Gamma$ point (Fig. 2(c)) in LaAgBi$_2$. The band $\alpha$ of LaAgBi$_2$ centered at $X$ point (Fig. 2(a)) which hosts Dirac fermions remains unaffected by the nesting or CDW transition. The band 3 in LaAgSb$_2$ separates to two bands ($\beta, \eta$) (Fig. 2(d)) in LaAgBi$_2$ which lose the nesting condition. So the bands which host Dirac fermions in LaAgBi$_2$ are most likely intact at the CDW transition.

\section{Conclusion}
In summary, first-principle calculation, de Hass-van Alphen oscillation study of the electronic structure and magnetoresistance behavior of LaAgBi$_2$ show striking similarity to properties of (Sr,Ca)MnBi$_2$. LaAgBi$_{2}$ has no magnetic ions and is a paramagnetic metal without long range magnetic order. Yet, the band structure clearly shows several narrow bands with nearly linear energy dispersion and Dirac-cone-like points at the Fermi level. This is in agreement with the quantum oscillation experiments that revealed three Fermi pockets with small cyclotron resonant mass. The in-plane transverse magnetoresistance exhibits a crossover at a critical field $B^*$ from semiclassical weak-field $B^2$ dependence to the high-field unsaturated linear magnetoresistance which is a hallmark of the quantum limit of the Dirac fermions. The temperature dependence of $B^*$ satisfies quadratic behavior, which is attributed to the splitting of linear energy dispersion in high field. Our results demonstrate the possible universal existence of two dimensional Dirac fermions in layered structure compounds with two-dimensional Bi square nets, irrespective of magnetic order.
\begin{acknowledgments}
We than John Warren for help with SEM measurements. Work at Brookhaven is supported by the U.S. DOE under contract No. DE-AC02-98CH10886. The high magnetic field studies in NHMFL were supported by NSF DMR-0654118, the State of Florida and DOE NNSA DE-FG52-10NA29659.
\end{acknowledgments}

% Create the reference section using BibTeX:
%\bibliography{basename of .bib file}

\end{document}